 \documentclass{emulateapj}

\usepackage{apjfonts}

\slugcomment{to appear in ApJ, 2011}

\shorttitle{Slow Nova Evolution with a Companion}
\shortauthors{Kato \& Hachisu}

\begin{document}


\title{Effects of a Companion Star on Slow Nova Outbursts -- Transition 
from Static to Wind Evolutions}


\author{Mariko Kato}
\affil{Department of Astronomy, Keio University, 
Hiyoshi 4-1-1, Kouhoku-ku, Yokohama 223-8521, Japan:}
\email{mariko@educ.cc.keio.ac.jp}

\and

\author{Izumi Hachisu}
\affil{Department of Earth Science and Astronomy, 
College of Arts and Sciences,
University of Tokyo, Komaba 3-8-1, Meguro-ku, Tokyo 153-8902, Japan;}
\email{hachisu@ea.c.u-tokyo.ac.jp}




\begin{abstract}
Two types of nova evolutions can be realized
in low-mass white dwarfs of $\sim 0.5-0.7~ M_\sun$, 
i.e., an evolution with optically thick winds like in usual 
classical novae, or an another type of 
evolution without them like in the symbiotic nova PU Vul. 
The latter type is characterized by spectra of no indication of strong winds 
as well as a long-lasted flat optical peak in its light curve. 
We propose a transition from no-optically thick wind evolution to 
usual evolution with optically thick winds as 
a new outburst model for slow novae that show a relatively long-lasted 
multipeak phase followed by a wind phase like in the slow novae 
V723 Cas, HR Del, and V5558 Sgr.
 We calculated nova envelopes with one-dimensional approximation of 
the companion's effects and  
found that when the companion star is deeply embedded in the extended nova 
envelope, the structure of static envelope 
approaches that of the optically thick wind solution. 
Thus, the transition from static to wind solution is triggered 
by the effect of the companion.  
The transition occurs in a close binary nova like  
V723 Cas, but is not triggered in a long period binary like 
PU Vul. 
We reconfirm our previous results that the frictional energy deposition is 
negligibly small in almost all of hydrogen/helium novae    
because of the low envelope density at the orbit. 
\end{abstract}





\section{Introduction} \label{sec_introduction}

Nova is a thermonuclear runaway event on a mass-accreting  white dwarf (WD).
During a nova outburst, the envelope on the WD expands considerably and 
engulfs its mass-donating companion unless it is a wide binary.
As the companion orbits in the nova envelope, frictional processes
between the envelope and the companion produces thermal energy. 
In 1980's, it was believed that these frictional processes play 
an important role to eject the envelope and  
to shorten the nova duration by a factor of 10 or more
\citep{mac85, liv90, sha91}. 
This idea was originally proposed to reconcile observed short durations  
of classical novae ($\sim$ 1 yr) with long timescales of nuclear burning
($\sim$ 1000 yr).  However, the drag energy deposition is not effective
for mass ejection in the common envelope phase of novae
because the envelope masses are too small 
to produce drag-energy enough to eject the envelope \citep{kat91a,kat91b}.

In the beginning of 1990's, opacity tables were revised 
\citep[OPAL opacity:][]{igl96}, of which a prominent peak at $\log T $(K)$ 
\sim 5.2$ is strong enough to accelerate optically thick winds
even in relatively less massive WDs of $\sim 0.6~M_{\sun}$. 
This wind is as massive as $10^{-4}~M_\sun$~yr$^{-1}$ or more. 
The nova duration is drastically reduced so that the theoretical 
duration becomes comparable to the observed one \citep[e.g.,][]{kat94h,pri95}. 
Once the wind occurs, frictional effects are ineffective, because the 
density at the orbit is too low to produce large energy 
deposition. Moreover, the wind is accelerated deep inside the orbit and the 
wind velocity has already reached the escape velocity at the orbit. 
Thus, once the optically thick winds occur, the presence of
a companion has virtually no effects on the envelope structure
nor on the duration of nova outbursts \citep{kat94h}.

There are observational indications that frictional process
is not effective in nova outbursts. 
If the frictional process is very effective in mass ejection, 
there should be a clear dependence of nova light curve shapes on 
the orbital period, but no such properties are known. 
For example, the recurrent nova RS Oph shows a very fast development of
optical light curve \citep{hac06b} similar to that of U Sco. RS Oph is
a long period binary  of $P_{\rm orb}=456$ days \citep{fek00}, so that
the WD photosphere does not reach the companion even at the maximum
expansion \citep{hac06b}. Therefore, the frictional processes of
companion do not work in the whole period of the nova outburst. 
On the other hand, U Sco is a short period binary of 
$P_{\rm orb}=1.23$ days \citep{sch95} and the companion star is
engulfed deeply in the envelope during the outburst.
The 1999 and 2010 outbursts of U Sco were densely observed,  
but there is no indication for additional acceleration nor enhanced 
mass ejection due to the companion motion \citep{hkkm00}.

We have analyzed a number of nova light curves 
with the universal decline law based on the optically thick winds 
\citep{hac06,hac07k,hac10,hac06b,hac08,kat94h,kat09,kat09b}.
If the companion motion strongly contributes to the acceleration of matter, 
there should be a change in the decline rates of light curves when the 
companion reappears from the extended envelope of the WD.
However, none of them show a particular change in the light curves 
before and after the epoch that the companion reappeared from the 
envelope. We refer two examples of classical novae, with a densely observed 
light curve and known orbital period: V1500 Cyg 
\citep[$P_{\rm orb}=0.14$ days,  
see the light curves in Figure 13 of][]{hac06}; the companion emerged
on days $\sim 50$ but nothing had happened,   
and also V458 Vul \citep[$P_{\rm orb}=0.59$ days,
in Figure 26 of][]{hac10}; the companion emerged on days $\sim 60$
but nothing happened.  These are good counterevidence against
the effectiveness of frictional process.  
Therefore, we emphasize that frictional processes are not effective 
in novae whenever strong optically thick winds blow.

In our theoretical models, optically thick winds always occur
in nova outbursts 
on massive WDs ($ \gtrsim~0.7~M_\sun$) but do not occur in less massive WDs 
($\lesssim~0.5~M_\sun$).  Between them, i.e., $\sim$0.5--0.7 $M_\sun$,
both types of nova evolutions, with/without optically thick winds,  
are realized for the same WD mass \citep{kat09}.
When no optically thick winds are accelerated, nova outbursts show a 
slow evolution with a long-lasted flat optical maximum, in contrast with usual 
classical novae that show a sharp optical maximum caused by a rapid evolution 
due to wind mass loss \citep{kat11s}.  
The first example of such no-optically thick wind evolution
is the symbiotic nova PU Vul, in which   
a flat optical peak lasted 3000 days and showed very quiet spectra  
indicating no strong winds \citep{kat11}.

\citet{kat09} presented an idea that a nova outburst started  
in a state of no-optically thick winds (we call this ``static evolution'')
and then possibly changes to be in a state of 
optically thick winds (``wind evolution'').
They also suggested that such a transition may accompany violent 
activities such as oscillatory behaviors in the light curves
of some slow novae.
 
Here, we present an idea that a companion star plays a role in 
changing the nova envelope structure from ``static'' to ``wind.'' 
If this occurs in low-mass WDs, it may relate to the peculiar light 
curves of slow novae.
As an application of such a transition to slow novae like V723 Cas,  
we have examined effects of a companion star with one-dimensional (1D)
approximation of nova envelopes. 
In Section \ref{sec_model}, we briefly introduce our method. 
Numerical results are presented in Section \ref{sec_results}. 
Applications to slow novae appear in Section \ref{sec_slownova}. 
Discussion and summary follow in Sections  \ref{sec_discussion}
and \ref{sec_summary}.

\section{Envelope Model}   \label{sec_model}
We consider low-mass WDs in which no optically thick winds are accelerated 
in the beginning of nova outbursts. This may happen in low-mass WDs 
($\lesssim 0.7 ~M_\sun$) under some conditions of the initial envelope 
mass and chemical composition \citep[see][for more details]{kat09}. 
After a thermonuclear runaway sets in, the envelope greatly  
expands and the optical brightness increases to reach a flat maximum. 
We approximate the nova evolution by a sequence of hydrostatic solutions 
when no optically thick winds occur, and by a sequence of steady-state 
solutions with optically thick winds when it occurs. 
In these sequences of solutions the envelope mass is decreasing 
due to nuclear burning and wind mass loss if it occurs,
from which we calculate evolution timescale. This quasi-evolution is 
a good approximation because the thermal/dynamical timescale between two successive 
solutions are short enough compared with the evolution timescale of novae. 
 
For static solutions  we solve the equations of hydrostatic balance,
mass continuity, radiative diffusion, and conservation of energy,
from the bottom of the hydrogen-rich envelope through the photosphere
assuming spherical symmetry. For wind solutions, we solve the equation of 
motion assuming steady-state, instead of hydrostatic balance \citep{kat94h}.
Convective energy transport is calculated in static solutions using 
the mixing length theory with the ratio of mixing length to pressure scale 
hight $\alpha=1.5$ \citep[see][for effects of the $\alpha$ parameter
on static solutions]{kat09}. The occurrence of optically thick winds is 
detected by the condition described in \citet{kat85}.
In our numerical calculations, we adopt more than 8000 meshes 
and always insert $\sim 1700$ meshes in the interacting 
region between $r=R_{\rm orb}-R_{\rm a}$ and $r=R_{\rm orb}+R_{\rm a}$, 
where $R_{\rm orb}$ is the position of the companion from the WD center and 
$R_{\rm a}$ is the accretion radius as defined below. 
The chemical composition of the envelope is simply assumed to be uniform with 
the composition of $X=0.55,~Y=0.23, ~CNO=0.2$, and $Z=0.02$, which are 
representative values for slow novae \citep[see Table 1 of][]{hac06}. 
We assume an 0.55 $M_\sun$ WD and an 0.4 $M_\sun$    
main-sequence (MS) companion that fills its Roche lobe unless otherwise stated. 
This 0.4 $M_\sun$ is close to an upper limit of the companion mass,
0.44 $M_\sun$, in a binary system with thermally stable mass transfer 
($q=M_{\rm comp}/M_{\rm WD} \lesssim 0.8$). We assume such a large mass because a 
more massive companion has larger effects on the WD envelope. 

Three effects of the companion on the WD envelope are considered here:  
(1) spun-up by the companion motion, 
(2) gravity of the companion star, and 
(3) drag luminosity due to frictional energy deposition.  
We incorporated these effects in our computer code as explained in the 
following subsections.

\subsection{Centrifugal Force} \label{centrifugal_force}

In the very beginning of a nova outburst, the envelope has so small specific
angular momentum compared with the specific orbital angular momentum.
After the envelope expands to a giant size, it begins to counterrotate 
on the rotating frame with the companion due to conservation of the local
angular momentum.  The outer part of the envelope may be spun up  
due to frictional effects of the companion motion. 
We incorporate the effect of centrifugal force in our 1D code. 
We modified the equation of hydrostatic balance as 
\begin{equation}
{1\over \rho}{dP\over{dr}}=-{{GM_r}\over r^2}+r\omega^2,
\label{equ_hydrostatic}\end{equation}
\noindent
where $\omega$ is the angular velocity.
This is a 1D approximation that may represent envelope 
structure in the equatorial plane. 
Here we introduce the effective mass parameter $f$ and rewrite 
Equation (\ref{equ_hydrostatic}) as 
\begin{equation}
{1\over \rho}{dP\over{dr}}=-{{GM_r}\over r^2} f,
\label{equ_hydro.f_def}
\end{equation}
\begin{equation}
f \equiv 
1-  {\omega^2~r^3 \over{GM_r}}.
\label{equ_f_def}
\end{equation}
The factor $f$ represents the degree how much the gravity
is effectively reduced due to the centrifugal force.
In other words, the effective mass decreases 
to $fM_r$. We, hereafter, call $f$ the mass reducing factor. 

We assumed a simplified rotation law for the envelope; a rigid rotation in the 
inner part of the envelope ($r < R_\Omega$) 
and a Keplerian-like rotation in the outer part ($r \geq R_\Omega$).
Since the envelope is spun up by the companion motion inside the 
outer Lagrange point, we define  the rigid rotation radius as
$R_\Omega=R_{\rm orb}+1.2 \times R_{\rm comp}$ 
because the outer Lagrange point is located at $\sim 1.2 \times R_{\rm comp}$ 
from the companion center.  

For the inner part ($r < R_\Omega$), we assume a solid rotation with 
an angular velocity of $\omega_0=\eta V_{\rm orb}/R_{\rm orb}$, 
here $\eta$ is a parameter that represents the ratio of 
the angular velocity of the envelope to that of the companion, 
$\Omega = V_{\rm orb}/R_{\rm orb}$. In other words, $\eta$ means 
the ratio of the envelope velocity to the companion velocity at the orbit.

For the outer part ($r \geq R_\Omega$) we assume that the rotation velocity 
of the envelope varies as $r^{-1/2}$ as the Keplerian velocity does.  
In this case, the angular velocity $\omega$ changes
as $r^{-3/2}$, and can be expressed as 
$\omega = \eta V_{\rm orb} R_{\rm orb}^{-1}R_\Omega^{3/2} r^{-3/2} $.  
Using Kepler's law $V_{\rm orb}^2 =
G (M_{\rm WD} + M_{\rm comp})/R_{\rm orb}$, 
the specific centrifugal force can be rewritten as 
\begin{equation}
r\omega^2  = 
\eta^2 ~{({R_\Omega \over R_{\rm orb}})}^3
 ~{{G({M_{\rm WD}+M_{\rm comp})}}\over r^2}.
\label{equ_centrifugal}
\end{equation}
Then $f$ becomes
\begin{equation}
 f = 1-\eta^2~({R_\Omega \over R_{\rm orb}} )^3,
 ~~~{\rm for~~}r \geq R_\Omega. 
\label{equ_f2}
\end{equation}
Here we use $M_r = M_{\rm WD}+M_{\rm comp}$ at $r \geq R_\Omega$. Thus,  
$f$ is constant at $r \geq R_\Omega$. 
Approximations of these centrifugal force and 1D approximation are discussed 
in Sections \ref{discussion_spherical} and \ref{discussion_rotation_law}.

\subsection{Gravity of the Secondary}

In our 1D approximation, we have simply 
assumed that the companion is distributed uniformly within a shell 
of $r=R_{\rm orb} \pm R_{\rm comp}$. 
Since the envelope mass ($10^{-3}$ to $10^{-6}~M_\sun$) is much smaller
than that of the companion ($M_{\rm comp}=0.4~M_\sun$),
the change of $M_r$ is essentially due to the mass 
distribution of the companion, i.e., $M_r= M_{\rm WD}$ inside the radius of 
$r=R_{\rm orb}-R_{\rm comp}$, and  $M_r = M_{\rm WD}+ M_{\rm comp}$ outside 
the radius of $r=R_{\rm orb}+R_{\rm comp}$. 
These treatments are essentially the same as those in \citet{kat94h}. 

\begin{figure}
\epsscale{1.15}
\plotone{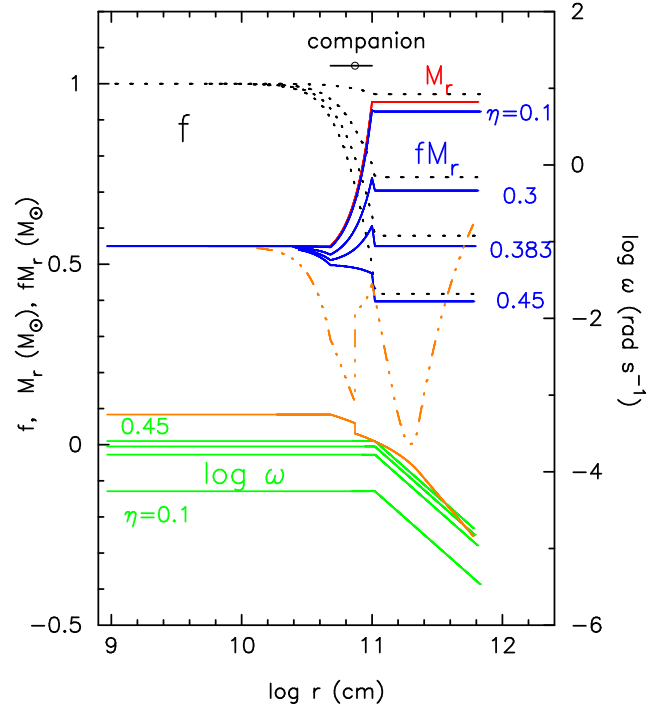}
\caption{
Adopted rotation law $\log \omega$ (green solid lines), 
gravity reducing factor $f$ (black dotted lines), 
mass distribution $M_r$ (red solid line), 
and effective mass, $f M_r$ (blue solid lines), for a binary consisting of 
an 0.55 $M_\sun$ WD and 
an 0.4 $M_\sun$ MS companion with $\eta$=0.1, 0.3, 
0.383, and 0.45, from upper to lower for $f$ and $f M_r$, but 
from lower to upper for $\log \omega$. The corresponding envelope structures 
are shown in Figure \ref{struc.compari}.
The left/right edge of each line corresponds to the bottom/photosphere 
of the envelope.  
The location of the companion ($\log r$ (cm)=10.869) and its size are 
indicated by a small open circle with a short horizontal bar.
The orange solid and dash-three-dotted lines denote a trial rotation law 
and the corresponding 
effective mass, respectively (see Section \ref{discussion_rotation_law}). 
\label{drag}}
\end{figure}

Figure \ref{drag} shows examples of 
the angular velocity $\omega$, mass distribution $M_{\rm r}$, 
mass reducing factor $f$, and effective mass $fM_{\rm r}$ 
for our envelope models. Here, 
$f$ is unity in the deep inside the envelope ($r \ll R_{\rm orb}$), but 
0.971, 0.741, 0.578 and 0.418 in the outer region ($r \geq R_\Omega$)  
for $\eta=0.1,~0.3,~ 0.383$, and 0.45, respectively.
When $\eta=0.383$ ($f=0.578$), we have  
$(0.55~M_\sun+0.4~M_\sun)\times 0.578= 0.55~M_\sun$ at $r \geq R_\Omega$, 
i.e.,  the companion gravity is almost canceled by 
the centrifugal force. 
A small peak of the effective mass 
appears near $\log r$ (cm)$ \sim$ 11.0 
because $M_r$ is constant (i.e., $M_{\rm WD}+ M_{\rm comp}$) 
outside $r = R_{\rm orb}+R_{\rm comp}$, whereas 
$\omega$ begins to decrease from a bit outside, at $R_\Omega = R_{\rm orb} 
+ 1.2 \times R_{\rm comp}$

\subsection{Drag Luminosity}

We treat the drag luminosity in spherical (1D) approximation 
in the same way as in the other previous works \citep{taa89,liv90,kat91a,kat91b,kat94h}.
The drag luminosity generated in the region from $r$ to 
$r+\delta r$ can be approximated as 
\begin{equation}
\delta L_{\rm drag} = \rho~ (V_{\rm orb}-v_{\rm e})^3 ~\delta S,
\end{equation}
\noindent
where $\rho$ is the density of the envelope, $V_{\rm orb}-v_{\rm e}$ is 
the relative velocity between the envelope and the companion star, 
$\delta S$ is the cross-sectional area between a circular strip from $r$ to $r+\delta r$ 
and a circle with the radius $R_{\rm a}$, the center of which is located 
at the orbit, $r=R_{\rm orb}$. 
Here the modified accretion radius, $R_{\rm a}$,
is defined as 

\begin{equation}
R_{\rm a} = {R_{\rm 0} \over {1+ (R_{\rm 0}/2H)^2}}, 
\label{equation_Ra}\end{equation}
where $H$ is the local density scale height, and $R_{\rm 0}$ 
is the generalized \citet{bon52} radius defined by 
\begin{equation}
R_{\rm 0} = {2 G M_{\rm comp} \over{[(V_{\rm orb}-v_{\rm e})^2
+ C_{\rm s}^2]}}, 
\label{equ_ro}\end{equation}
\noindent
here $C_{\rm s}$ is the sound speed. 
In our 1D calculation the drag luminosity, $\delta L_{\rm drag}$,  
is re-distributed over the entire spherical shell between $r$
and $r+\delta r$.  The total drag luminosity is
\begin{equation}
L_{\rm drag} = \int\rho~ (V_{\rm orb}-v_{\rm e})^3 ~\delta S, 
\label{equ_drag}\end{equation}
\noindent
where the integral region is from $R_{\rm orb}-R_{\rm a}$ to 
$R_{\rm orb}+R_{\rm a}$. 
We take the companion radius itself, $R_{\rm comp}$,  as the accretion radius 
instead of  $R_{\rm a}$ 
when $R_{\rm a}$ is smaller than the companion radius.
If both the density and $R_{\rm a}$ are constant in space, the 
drag  luminosity becomes the standard expression, i.e.,
\begin{equation}
L_{\rm drag} = \pi R_{\rm a}^2  \rho (V_{\rm orb}-v_{\rm e})^3   
\label{equ_drag_simple}\end{equation}
\noindent
\citep{taa89,liv90}.

\section{Nova Envelopes with a Companion Star} \label{sec_results}

\subsection{Structure of Static Envelopes} 

\begin{figure}
\epsscale{1.15}
\plotone{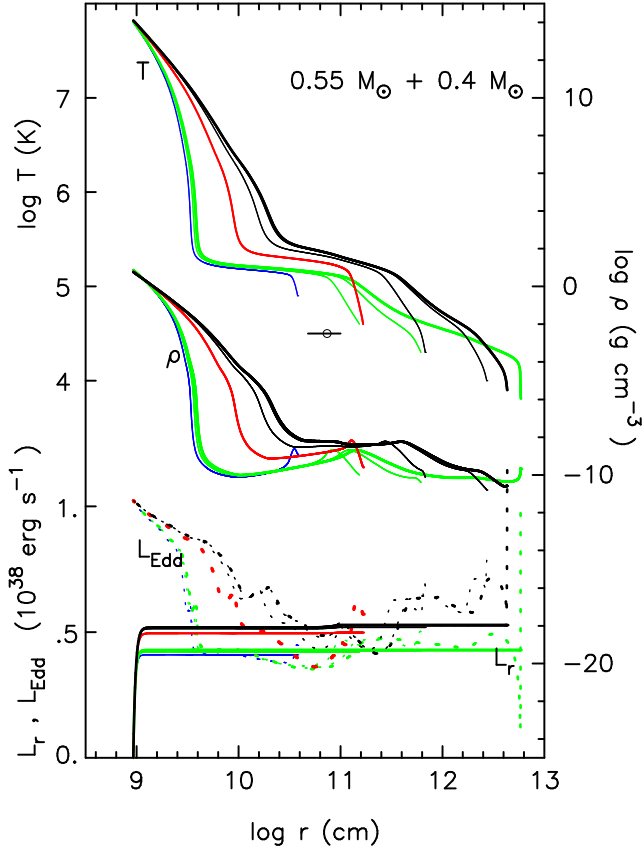}
\caption{
Distributions of the temperature $T$, density $\rho$, diffusive luminosity  
$L_{\rm r}$, and the Eddington luminosity $L_{\rm Edd}$ defined 
by Equation (\ref{equ_Edd}). Each line corresponds to the envelope 
with the photospheric temperature of $\log T_{\rm ph} {\rm (K)}=3.9$ (thick black lines),
4.0 (middle thin black lines), 4.3 (left thin black lines), and 4.6 (red lines).
The location of the companion 
and its size are indicated by the small open circle with a short horizontal bar.
For comparison, four solutions without the companion effects are added:  
$\log T_{\rm ph} {\rm (K)}=3.81 $ (green thick lines: right) and 4.3, 4.6 
(green thin lines: right and left, respectively), and 4.9 (blue thin lines).
The two envelopes of thick lines (with/without the companion effects)  
have the same envelope mass of $4.1\times 10^{-5}~M_\sun$. 
\label{struc.norotation}}
\end{figure}

Figure \ref{struc.norotation} shows 
distribution of the temperature, density, diffusive luminosity, 
and the local Eddington luminosity. 
We do not include the effects of a companion for blue and green lines. 
The local Eddington luminosity (dotted lines) is defined as
\begin{equation}
L_{\rm Edd} = {4\pi cGM_r \over\kappa},
 \label{equ_Edd}\end{equation}
where $\kappa$ is the opacity. We used the OPAL opacity \citep{igl96}.
As the opacity is a function of the temperature and density, 
and $M_r$ is also the function of the radius (see Figure \ref{drag}),  
the Eddington luminosity is also a local variable. 

The blue and three green lines in Figure \ref{struc.norotation} depict
the envelope solutions of $\log T_{\rm ph}$ (K)=4.9, 4.6, 4.3 and 3.81
for the case of no companion star.   These envelopes have essentially
the same structure except for the surface region.
As reported by \citet{kat09}, a static envelope has a core-halo 
structure of the density in which a large density inversion 
layer develops deep inside the photosphere. 
The density-inversion region corresponds to the super-Eddington region.
The local Eddington luminosity has the deepest local minimum
between $\log r$ (cm) $\sim$ 10.0 and 11.0, 
corresponding to the peak of the OPAL opacity at $\log T$ (K) $\sim 5.2$.
This density-inversion arises in order to keep hydrostatic balance 
in the super-Eddington region ($ L_r > L_{\rm Edd} $) as expected 
from the equation of hydrostatic balance \citep{kat09}. 
Inefficient convections occur in the region of  $L_r > L_{\rm Edd}$ but are 
unable to carry all of the diffusive energy flux.

Black and red lines in Figure \ref{struc.norotation} 
indicate the models in which the companion's gravity 
and drag luminosity are incorporated. The three envelopes denoted by the 
black lines have very similar structures at $\log r$ (cm) $ < 11.5$. 
The drag energy deposition is small as explained below.
Therefore, the diffusive luminosity barely increases around the orbit in 
the model of $\log T_{\rm ph}$ (K)=3.9, in which the drag luminosity
deposition is 1.8\% of the total flux. 

Comparing these envelope models (black lines) with those without the
companion effects (green and blue lines), we see that the envelope matter
is significantly re-distributed due to the companion's effects. In the
presence of the companion, the Eddington luminosity, outside the orbit is 
$L_{\rm Edd} = 4\pi cG(M_{\rm WD}+M_{\rm comp})/\kappa$, which is larger
than that without companion $L_{\rm Edd} = 4\pi cGM_{\rm WD}/\kappa$. 
This difference causes different energy flux as shown in Figure 
\ref{struc.norotation}, $5.3 \times 10^{37}$ erg~s$^{-1}$ in the model
of black thick lines, whereas $4.3 \times 10^{37}$ erg~s$^{-1}$ in the
green thick lines.  Therefore, these two envelopes are in different
hydrostatic balance.  A lower luminosity leads to a larger local 
super-Eddington region which causes a wider density-inversion region.
The solution depicted by red lines has a photosphere just outside
the companion orbit. Its photospheric luminosity is close to those of the solutions
with the companion, but the structure is something between the two
solutions with/without the companion.

\begin{figure}
\epsscale{1.15}
\plotone{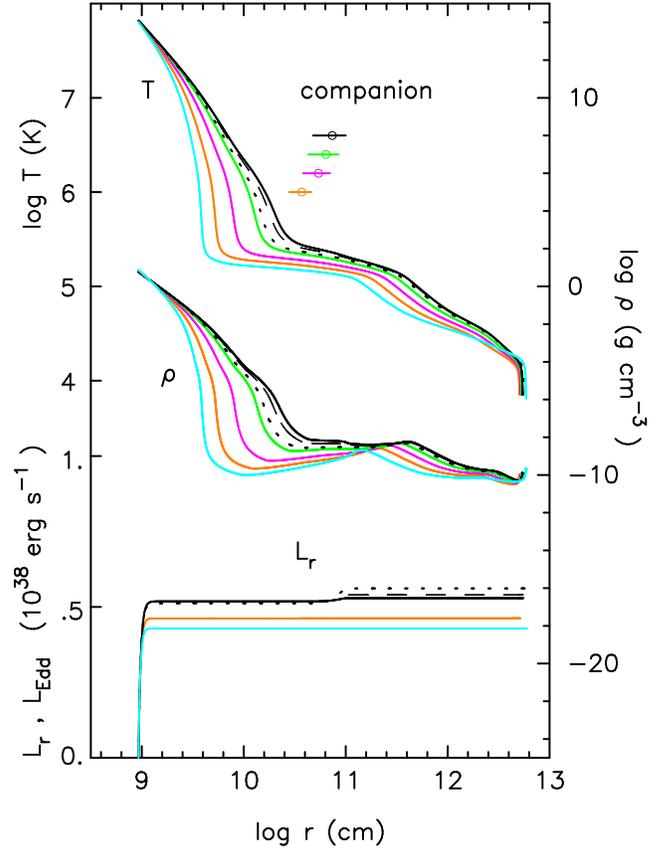}
\caption{
Same as Figure \ref{struc.norotation} but for static solutions with 
different companion masses. The photospheric temperature is $\log T_{\rm ph}$ (K)= 3.85 
for all the solutions. $M_{\rm comp}$= 0.0 (blue), 0.1 (brown), 0.2 (pink), 0.3 (green), 
and 0.4 $M_\sun$ (black solid). Diffusive luminosity for $M_{\rm comp}$= 0.2 and 0.3 
$M_\sun$ models are omitted to simplify the figure. 
The location of the companion are indicated by the small open circles with a 
short horizontal bar for $M_{\rm comp}$= 0.1, 0.2, 0.3, and 0.4 $M_\sun$ from left to 
right.  
Dashed and dotted lines indicate static solutions with enhanced 
drag luminosity by a factor of 2.8 and 10, respectively (see 
Section \ref{discussion_spherical} for details).
\label{struc.Mcomp}}
\end{figure}

Figure \ref{struc.Mcomp} compares structures of envelope models for 
$M_{\rm WD}= 0.55~M_\sun$ among companion masses of $M_{\rm comp}=$
0.0, 0.1, 0.2, 0.3, and 0.4 $M_\sun$. For a larger companion mass,
the density inversion layer is smaller. In this way, the companion
gravity has a role of redistribution of the envelope matter 
through different hydrostatic balance of gravity and pressure-gradient;
a large luminosity causes larger radiation pressure gradient so that
the matter is pushed outward.   Dotted and dashed lines indicate the models
with enhanced drag luminosities which 
will be explained in Section \ref{discussion_spherical}.

When the envelope rotates, the centrifugal force has an effect to 
reduce the gravity by the factor $f$. 
Figure \ref{struc.compari} shows the envelope structure for $\log T_{\rm ph}$ (K)=4.3  
with different angular velocity of $\eta$=0.0, 0.1, 0.3, 0.383, and 0.45.
For a larger $\eta$, the centrifugal force becomes larger, 
which reduces the effective gravity.
When $\eta=0.383$, the gravity of the companion
($M_{\rm comp}=0.4~M_\sun$) is almost 
canceled as discussed earlier (see Figure \ref{drag}), 
and the envelope has a structure very close to that of the solution 
without the companion effects. 
If we further increase $\eta$ to 0.45, the envelope structure does not change much 
anymore except for the outer region where the temperature and density quickly drops 
(see the purple dash-dotted lines).

\subsection{Structure of Wind Solutions} \label{sec_wind}

\begin{figure}
\epsscale{1.15}
\plotone{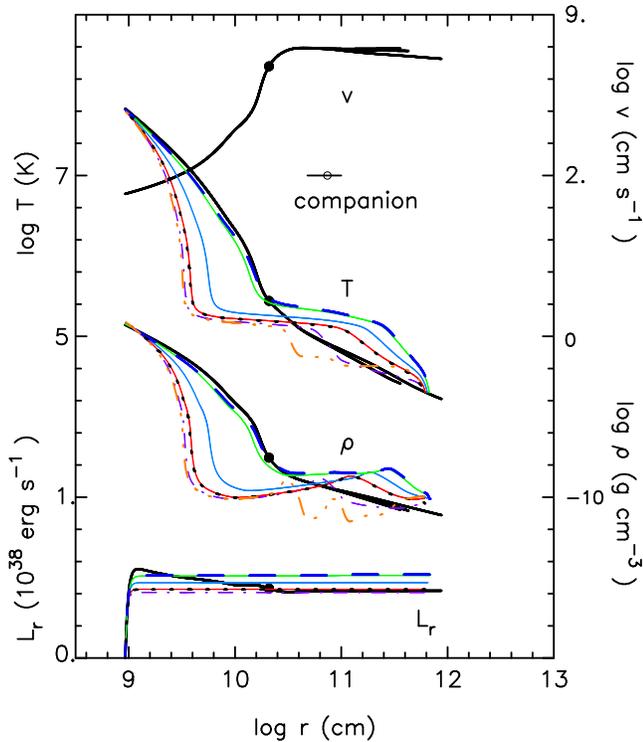}
\caption{
Comparison of the static and optically thick wind solutions with 
different rotation effect for the envelopes on 
a $0.55~M_\sun$ WD with a $0.4~M_\sun$ MS companion. 
Static solutions are depicted by colored lines, from upper to lower, 
$\eta=$0 (blue dashed line), 0.1 (green solid line), 0.3 (blue solid line), 
0.383 (red solid line), and 0.45 (purple dash-dotted line). 
Dotted black lines denote a static solution without the companion effects, 
which is almost 
overlapped with the static solution of $f=0.383$ (red solid line). 
All the static solutions have a photospheric temperature of 
$\log T_{\rm ph}$ (K) $=4.3$. 
Optically thick wind solutions are depicted by the black solid line with 
a black dot that indicates the critical point at $R_{\rm cr} = 0.3~R_\sun$. 
Three solutions of  $\eta=0.3, 0.383$, and 0.45 (the lower line
in $v$, but the upper lines in $T$ and $\rho$)
are almost overlapped with each other
as well as overlapped with the static solution of
$\eta=$0 (blue dashed line) inside the critical point.   
The orange dash-three-dotted lines denote a test solution with different 
rotation law (same as orange lines in Figure 1) which will be 
discussed in Section \ref{discussion_rotation_law}.
\label{struc.compari}}
\end{figure}

Figure \ref{struc.compari} also shows the structures of optically thick wind solutions. 
The optically thick winds are accelerated due to the peak of OPAL opacity 
which locates deep inside the photosphere. The wind velocity quickly 
increases around the critical point and reaches the terminal 
velocity deep inside the photosphere \citep{kat94h}. 

Optically thick wind solutions hardly change its structure 
once it is accelerated even if there are some changes of acceleration source like opacity  
in the outside of the critical point. 
Figure \ref{struc.compari} shows optically thick wind solutions for three 
different $\eta$, which correspond to different effective gravities as in 
Figure \ref{drag}.
As the critical point locates far inside the companion orbit  
and the winds are already accelerated there,  
the companion's gravity hardly changes the internal structure.

\subsection{Drag Energy Deposition during Nova Outbursts}\label{sec_drag}

We already found that the drag energy deposition is small
compared with the total luminosity.  For example, 
it is only 3.2\% of the photospheric luminosity in the model of 
$\log T_{\rm ph}$ (K) $=3.85$ 
(black solid line in Figure \ref{struc.Mcomp}). 
This small contribution comes from the low density near the companion orbit  
because the drag luminosity is proportional to the density. Using Equation 
(\ref{equ_drag_simple}), we estimate it as
\begin{eqnarray} 
L_{\rm drag} & \sim & \pi R_{\rm a}^2 \rho_{\rm orb} V_{\rm orb}^3 \cr
& = & \pi\times (3.3\times 10^{10} {\rm cm})^2 \times
 6.5\times 10^{-9}~{\rm g~cm}^{-3} \times (410~{\rm km~s}^{-1})^3  \cr
& = & 1.6 \times 10^{36}~{\rm erg~s}^{-1} \cr
& = & 0.030~L_{\rm ph},
\end{eqnarray}
almost the same as our model value of 3.2\%. 

This low density is caused by the opacity peak as shown in 
Figure \ref{struc.norotation}. The density is almost constant around 
the companion orbit. Thus, the drag luminosity is almost independent of
the orbital size. If we assume a less massive companion, 
then we have a smaller $R_{\rm a}$.  For a smaller companion mass,
on the other hand, the companion's Roche radius is smaller, so the 
companion comes into closer to the WD and the envelope's density is higher
at the companion's orbit.  However, a smaller radius (accretion radius)
effect overcomes the effect of slightly higher density.  As a result, a
$0.1~M_\sun$ companion produces a hundred times smaller drag luminosity
than an $0.4~M_\sun$ companion does.  Thus, an 0.4 $M_\sun$
companion gives almost an upper limit of the drag energy deposition
for the 0.55 $M_\sun$ WD model.  During a nova outburst,
the photospheric radius decreases with time and the density around
the orbit slightly decreases with time as shown
in Figure \ref{struc.norotation} in the case of static envelopes. 
Therefore, the drag luminosity decreases with time.   
In the case of rotating static envelopes, the density around 
the orbit is as small as or smaller than that
for the non-rotating envelope as shown in Figure \ref{struc.compari},
so the drag luminosity is also too small. 

In the case of optically thick wind envelopes, the density around the orbit 
is also as small as or 
much smaller than those of the static envelopes as shown in Figure \ref{struc.compari}. 
During the outburst, the density around the orbit decreases 
with time as will be shown later in Figure \ref{density.evolution}.
In this way, we do not expect large drag luminosities during nova outbursts 
either in the static or in the wind evolution.

\begin{figure}
\epsscale{1.15}
\plotone{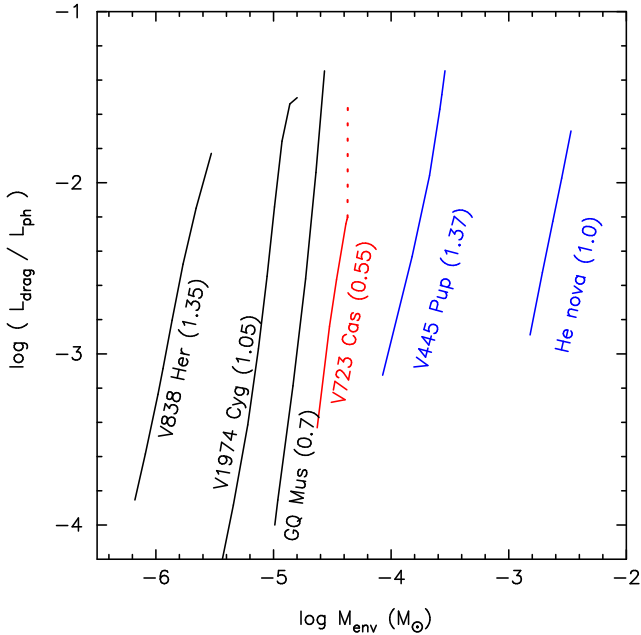}
\caption{
Ratio of the drag luminosity to the photospheric luminosity 
during nova outbursts. The top of each curve corresponds to the 
optical peak. The ratio decreases with time. The bottom point corresponds to
the stage that the companion emerges from the WD photosphere. 
The WD mass of each outburst model is indicated after the object name 
in solar mass units. 
Three black lines depict classical novae with optically thick winds: 
V838 Her (1.35 $M_\sun$), V1974 Cyg (1.05 $M_\sun$), and GQ Mus (0.7 $M_\sun$). 
A red line is the composite model for V723 Cas (0.55 $M_\sun$), consisting 
of a static phase 
(top dotted part) and a wind phase (bottom solid part).
The dotted part of V723~Cas represents the 
transition from the static to wind phase. Blue lines depict helium novae:    
V445 Pup (1.37 $M_\sun$) and a model for a 1.0 $M_\sun$ WD. 
See text for more details. 
\label{drag_efficiency}
}
\end{figure}

Figure \ref{drag_efficiency} shows the 
ratio of the drag luminosity to the photospheric luminosity estimated for 
individual nova outbursts. 
A nova outburst starts from the uppermost point 
of each line, which corresponds to the peak luminosity,  
and moves downward as the envelope mass decreases with time. The drag luminosity 
quickly decreases with time because the density around the orbit quickly decreases 
in wind solutions (see Figure \ref{density.evolution}).
The black lines depict three cases of classical novae. V838 Her is one 
of the fastest classical novae, which occurred on a very massive WD, thus the envelope 
mass is very small. Here we adopt the envelope structure taken from the 
light curve model of a 1.35 $M_\sun$ WD calculated by
\citet[][model 2 in their Table 3]{kat09b} and a   
temporarily adopted 0.45 $M_\sun$ companion at the separation
of $a=1.4 ~R_\sun$. 
The lowest point of the line corresponds to the stage at which the companion 
emerges from the photosphere and the companion effects 
vanish after that. 
In the same way, we have estimated the drag luminosities for the moderately fast 
nova V1974 Cyg \citep[a 1.05 $M_\sun$ WD model taken from][]{hac06}
with an 0.2 $M_\sun$ companion at $a=0.85~R_\sun$,   
and for one of the slowest classical novae GQ Mus
\citep[an 0.7 $M_\sun$ WD model taken from][]{hac08}
with an 0.1 $M_\sun$ MS companion at $a=0.6 ~R_\sun$. 
Here we adopt these parameters of the secondary, assuming 
they are in zero-age main sequence, which may not be accurate for each object.  
It is, however, enough for our purpose in estimating the drag luminosity, 
because it is not much affected by the choice of the secondary.

V838 Her and GQ Mus are the fastest and slowest classical novae, respectively. Thus, 
many other classical novae may fall in between these two lines like V1974 Cyg. 
We realize that the drag energy deposition is quite small. 
Even in the very early phase 
of the outburst, the ratio reaches only a few to several percent. 
Therefore, we reconfirm that the drag energy deposition does not 
contribute to the luminosity in classical novae.

Figure \ref{drag_efficiency} also shows the case of a ``transition nova,'' which is 
a model for V723 Cas (see the next section for more details). 
The binary consists of an $0.55~M_\sun$  WD and an  
0.4 $M_\sun$ MS companion.  The contribution of drag luminosity is also very small 
throughout the outburst.
  
The blue lines depict the case of helium nova outbursts. V445 Pup is only 
the known helium nova for which we adopt a $1.37~M_\sun$ WD model 
from \citet{kat08}. The companion is assumed to be a 1.0 $M_\sun$ helium 
star and $a=2.3~R_\sun$. 
For comparison, we also added another case of helium nova of
a 1.0  $M_\sun$ WD \citep[model taken from][]{kat08},
which has no observational counterpart but it 
is a theoretical purpose.  For the companion, we assume
an 0.8 $M_\sun$ helium star, and $a=2~R_\sun$. 
As shown in Figure \ref{drag_efficiency} the drag luminosity
is still very small.

\section{Application to  Slow Novae} \label{sec_slownova}
\subsection{Transition from Static to Wind Evolutions}

\citet{kat09} pointed out that two different types of nova evolutions, i.e., 
one is the evolution with optically thick winds and the other is 
without, can be realized 
in slow novae of a certain range of WD masses, $\sim 0.5-0.7~M_\sun$. 
For example, evolution of GQ Mus is explained by the sequence of 
optically thick wind solutions on a $\sim 0.7~M_\sun$ WD \citep{hac08}    
while the evolution of PU Vul is described by a sequence of hydrostatic solutions on 
a $\sim 0.6~ M_\sun$ WD \citep{kat11}. 

A remarkable difference between these two evolutions appears in 
the optical light curve \citep{kat11s}.  
The wind-type novae have a sharp peak 
in the optical light curve, because   
massive optically thick wind carries away a large part of the envelope 
mass and the nova light curves decay quickly. 
On the other hand, no-optically thick wind novae  
have a long-lasted flat optical peak before the magnitude slowly decays,  
because the nova evolves very 
slowly and it stays at an extended low-temperature stage for a   
long time, which makes a long-lasted flat optical peak.

\citet{kat09} presented an idea that a transition from the static evolution to 
the wind evolution could occur during an outburst. 
In such a case, the nova shows a flat optical peak with no indication 
of strong mass loss in the early phase of the outburst, 
followed by the decay like in normal novae with strong winds. 
\citet{kat09} further suggested that such a transition accompanies  
some activities like oscillatory behavior in brightness, because 
the internal structures of static/wind solutions are very different and the relaxation 
process may cause some oscillatory features.

Such a transition, however, may need some triggering mechanism. 
PU Vul has a flat optical peak that lasted for 8 yr. This indicates that  
the static evolution is stable in a timescale of $\sim 10$ yr. 
We suppose that a transition to a wind evolution was not triggered in PU Vul because 
the structures of the static envelope are very different from those of 
the wind solutions. 
The static solution, however, changes its structure 
when a companion star disturbs the structure of a static envelope 
as shown in Section \ref{sec_results}.  
This encourages us to further investigate our idea 
that the transition from the static to wind solution 
occurred in some slow novae.

\subsection{Comparison of V723 Cas, HR Del, and V5558 Sgr with PU Vul}

\begin{figure}
\epsscale{1.15}
\plotone{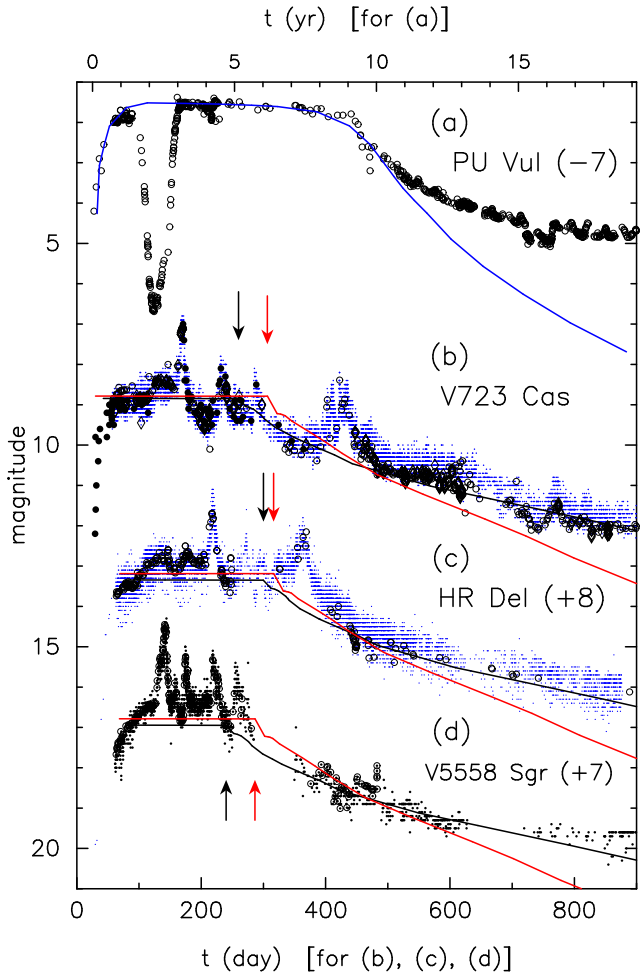}
\caption{
Comparison of light curves among PU Vul, V723 Cas, HR Del, and V5558 Sgr. 
PU Vul is shifted upward by 7 mag and HR Del and V5558 Cyg downward
by 8 and 7 mag, respectively.  
The upper timescale is for PU Vul and lower one is for the other three novae. 
(a) PU Vul: data taken from \citet{kat11}.
The dips at $t=2.2$ yr and 15.7 yr are eclipses.
(b) V723 Cas: diamonds \citep{cho97}, filled small circles (IAUC Nos. 
6213, 6214, 6227, 6233, 6256, 6283, 6331, 6358, 6428), 
open circles (AAVSO,${\it V}$-mag), dots (AAVSO, visual). 
(c) HR Del: open circles \citep[][${\it V}$]{sto67, poh67, nha67, ond68, oco68,
gry69, mol69, man70, bar70},  dots (AAVSO, visual). 
(d) V5558 Sgr: open circles (AAVSO, ${\it V}$), dots (AAVSO, visual).
The solid lines indicate the composite light curve model of 
$0.55~M_\sun$ WD with $X=0.55$, $C+O=0.2$, $Z=0.02$ (red line),  
and $0.6~M_\sun$ WD with the solar composition (black line).
The arrows indicate the switching point from a static to a 
wind evolution: $\log T$ (K)=3.88 (red) and 3.93 (black).
\label{4nova}}
\end{figure}

Figure \ref{4nova} shows the observational light curve of PU Vul,
V723 Cas, HR Del, and V5558 Sgr.   
PU Vul showed a flat optical maximum that lasted for 8 yr, 
neglecting the deep eclipse at $t=2.2$ yr. 
The light curves of V723 Cas, HR Del, and V5558 Sgr are very similar to 
each other as already pointed out by many authors
\citep{fri92,fri02,eva03,mun07}.
These novae have multiple peaks, showing an oscillatory variation 
around a certain magnitude. After that the magnitudes declined
rather smoothly. 

Resemblances in spectral evolutions were also suggested. 
V723 Cas shows very narrow emission/absorption lines in the spectra
at the pre-maximum flat peak. The spectrum gradually changes to 
a pure absorption F-type supergiant before the optical maximum
\citep{iij98}. Note that PU Vul also shows F-type supergiant spectra
with pure narrow absorptions, which are interpreted as static evolution
\citep{kat11}.  After the optical maximum, V723 Cas changes its spectrum
to that of usual novae \citep{iij98} with many emission lines, some of
which show a P Cygni profile \citep{iij06,eva03}. 
After that it entered the nebular phase (645-677 days after the discovery). 

The spectral evolution of V5558 Sgr before and after the maximum is also  
very similar to that of V723 Cas \citep{iij07b,mun07,pog08,pog10}. 
In the pre-maximum phase the nova shows a featureless F-type spectrum except 
narrow emission lines H$\alpha$ and H$\beta$ and many weak absorption lines 
\citep{iij07a,nai07,tan11}, which does not resemble to those
of classical novae. 
Fast hot winds (FWHM 1150 --1500 km~s$^{-1}$)   
appeared in the multi-peak episode suggest that mass ejection is 
associated to the brightening \citep{mun07,kis07,tan11}. 
During the decline phase, V5558 Sgr had entered the nebular phase and the 
spectra resemble to those of V723 Cas \citep{pog10}.

HR Del also showed spectral evolution unlike classical novae 
\citep{hut70,san74,raf78}.
The early spectra before maximum were characterized by relatively narrow
emission lines with P Cygni absorption, and the spectral class of F0. 
\citet{fri92} examined spectral evolution before maximum and concluded that 
optically thick winds unlikely occurred before maximum and that the 
photosphere is almost stationary. 
HR Del showed several optical maxima which accompany discrete shell
ejections \citep{raf78,san74}. 
In the nebular phase, spectral evolution was alike to normal novae
\citep{raf78}.  
 
These spectral changes in V723 Cas, V5558 Sgr, and HR Del are 
consistent with our suggestion that these novae 
started its outburst from no optically thick wind evolution  
and, after the multi-peak phase,  
it changes into an optically thick wind evolution like usual novae. 
%

\begin{deluxetable*}{lllllllll}
\tabletypesize{\scriptsize}
\tablecaption{Table 1. Slow Novae with Long-lasted Peak/Pre-maximum Phase
\label{table_nova_parameters}}
\tablewidth{0pt}
\tablehead{
\colhead{Object} &
\colhead{} &
\colhead{Outburst} &
\colhead{Maximum Phase} &
\colhead{Multi-peak} &
\colhead{Dust\tablenotemark{a}} &
\colhead{$P_{\rm orb}$} &
\colhead{$M_{\rm WD}$\tablenotemark{b}} &
\colhead{Companion}
}
\startdata
PU Vul       &... &1979& 8 yr  &no & no & 4900 days &$\sim 0.6~M_\sun$& M giant \\
V723 Cas & ... & 1995& 400 days  & yes& no& 16.6 hr  &$0.58, 0.59~M_\sun$& dwarf \\
HR Del  & ...  & 1967& 320 days & yes& no &5.14 hr  &0.595, $\sim 0.9~M_\sun$  &dwarf   \\
V5558 Sgr & ... & 2007& 200 days &yes & no& unknown &$0.58-0.63~M_\sun$  &dwarf\tablenotemark{c}
\enddata
\tablenotetext{a}{PU Vul: there was a debate on the origin of the deep minimum, i.e., 
to be dust origin or eclipse, but later, it turned clear to be an eclipse.  
V723 Cas: no dust formed \citep{lyn00}. 
HR Del: no indication of dust in the spectral evolution \citep{raf78}. 
V5558 Sgr: no dust formed \citep{rud11}.
}
\tablenotetext{b}{PU Vul: about 0.6 $M_\sun$ by \citet{kat11} from light curve fitting, 
  V723 Cas: 0.58 $\pm 0.07~M_\sun$ by \citet{iij98} from absolute B magnitude; 
   0.59 $M_\sun$ by \citet{hac04} from light curve fitting. 
  HR Del: 0.595 $M_\sun$ by \citet{kue88} from radial velocities; 
   0.9  $M_\sun$ by \citet{bru82} from radial velocities. 
  V5558 Sgr: 0.58-0.63 $M_\sun$ by \citet{pog10} from relation between absolute magnitude 
  at maximum and the WD mass. 
}
\tablenotetext{c}{see the text.}
\end{deluxetable*}

Table \ref{table_nova_parameters} summarizes the outburst properties
of these four novae, i.e., the outburst year, duration of maximum phase, 
presence of multi-peak, information on dust formation, orbital period,
estimated WD mass, and suggested type of the companion. 
For PU Vul, the maximum phase means the phase of flat 
maximum. For the other three novae, we define the 'maximum phase' 
by the period between the end of quickly rising phase and the last
prominent peak.  
We see that these `maximum phases' last 200-400 days, which is much shorter 
than the flat maximum phase of PU Vul (8 yr), but much longer than the orbital 
periods, so there is sufficient time for the companion to orbit many
times around the WD before the envelope makes the transition. 

PU Vul is a long period binary of $P_{\rm orb}=4900$ days and 
the companion is an M type red giant (RG). 
During the nova outburst, the WD envelope expands to $\sim 60~R_\sun$, which 
is much smaller than the companion orbit \citep{kat11}. 
Thus, the envelope structure is not affected by the companion.
On the other hand, V723 Cas \citep[$P_{\rm orb}=0.693$ days = 16.6 hr:
][]{gor00} and HR Del \citep[$P_{\rm orb}= 0.214$ days =5.14 hr:
][and references therein] {fri10,kue88} are close binaries and
the companion is a dwarf. 
The orbital period of V5558 Sgr is unknown, but suggested to be as
short as normal classical novae, because a large outburst amplitude
\citep[$\Delta V > 10$ mag: ][]{nai07} suggests that the companion is not
an RG.
Therefore, in these three novae, the companion should have been deeply
embedded in the nova envelope during the outburst.  

This is, again, very consistent with that PU Vul (a wide binary) 
underwent a quiet evolution without strong winds throughout the outburst 
\citep{kat11}. No transition occurred in PU Vul. On the other hand, V723 Cas, HR Del, and 
V5558 Sgr (close binaries) outbursted as a quiet evolution at the beginning, 
then changed into a wind evolution. 
The multiple peaks of these three nova light curves  
may be a relaxation process associated with the transition. 
The absence of multiple peaks in PU Vul suggests no transition triggered, 
which may also support our idea.

No indication of dust formation has been reported for all of the four novae
as in Table \ref{table_nova_parameters}. This is consistent with our
suggestion that no/weak optically thick winds occur in these novae,
because dust is usually formed in massive winds like OS And and DQ Her.

The estimated WD mass is about 0.6 $M_\sun$ for all three novae, V723 Cas, HR Del, 
and V5558 Sgr. Note that some works on slow novae mentioned that this  
 0.6 $M_\sun$ is a mass close to the lower limit of WDs 
having strong nova outbursts as a thermonuclear runaway event \citep{fri92,tkat02,tan11}. 
This argument is based on the calculation by \citet{kov85} with the old opacity. 
Recent calculations with the OPAL opacity, however, show that thermonuclear runaway 
triggers nova outbursts even in much less massive WDs, 
e.g., in 0.4 and 0.6 $M_\sun$ WDs \citep{sha93},
0.4 $M_\sun$ WD \citep{yar05}, and 0.2 - 0.5  $M_\sun$ WDs \citep{she09}. 
Therefore, $\sim 0.6~ M_\sun$ is not the lower mass limit
for a WD  having nova outbursts. 
Instead, we emphasize that 
$\sim 0.6~ M_\sun$ is a lower boundary for a nova outburst having  
optically thick winds for solar composition \citep{kat09}. 
Optically thick winds are weak,  
so it looks like a 'borderline classical nova' \citep{fri02}.

\subsection{Composite Light Curves}

Figure \ref{4nova}(a) shows theoretical light curves of PU Vul taken from 
\citet{kat11}, which consist of only static solutions.
The theoretical curve well 
represents the observational light curve until $t \sim 11$ yr,
and then deviates after that 
when the nova enters the nebular phase. This theoretical curve represents 
blackbody emission from the WD photosphere and 
does not include the contribution from nebular emission lines
outside the photosphere as well 
as of the RG companion which dominates in the later phase. 
Therefore, the theoretical magnitude is much lower  than the observation.  

We made composite theoretical light curves which mimic the light curves of 
V723 Cas, HR Del, and V5558 Sgr. Here, we assume that the early phase is 
approximated by a sequence of static solutions and the later phase 
by a sequence of optically thick wind solutions. 
We made two trial light curves; one is the model of 0.55 $M_\sun$ WD with 
the chemical composition of $X=0.55,~Y=0.23,~CNO=0.2$, and $Z=0.02$, and 
the other is an 0.6 $M_\sun$ WD with the solar composition $X=0.7,~Y=0.28,$ 
and $Z=0.02$.  We switched the solution from static to wind
when the photospheric temperature increases to $\log T_{\rm ph}$ (K)=3.89
for the 0.55 $M_\sun$ WD or to $\log T_{\rm ph}$ (K)=3.87 for the 
0.6 $M_\sun$ WD model.  The effect of the companion is included 
in the static phase assuming an 0.4 $M_\sun$ MS,
but not included in the wind phase for simplicity because of 
their small effects. No rotation effects are included. 

Our trial light curves are depicted in Figure \ref{4nova} (b)-(d).
Red lines indicate the model of 0.55 $M_\sun$ WD and black lines the model of 
0.6 $M_\sun$ WD. We applied the above two trial light curves to the three
novae, V723 Cas, HR Del, and V5558 Sgr.  The switching epoch from
a static to wind evolution is indicated by an arrow. 
Here, we assumed that the transition occurred $\sim 200$ days after 
the outburst. In the very beginning of outbursts, the transition unlikely occurs  
because the envelope structure may have not yet been close enough to that of a 
wind solution (see the red line solution in Figure \ref{struc.norotation}).   
Once the envelope has extended to a giant size, it may take several dynamical 
timescale for the structure change from static to wind solution. The dynamical 
timescale is estimated to be $t_{\rm dyn} \sim (2/GM)^{1/2} r^{3/2} = 
21$ days for $r=6\times 
10^{12}$ cm and $M=0.95~M_\sun$. Note this timescale is roughly consistent 
with the magnitude variation of these novae which may associate with 
a relaxation process. 
Apart from these oscillatory behaviors, our trial light curves are roughly 
consistent with the observed light curves of these three novae.

\begin{figure}
\epsscale{1.15}
\plotone{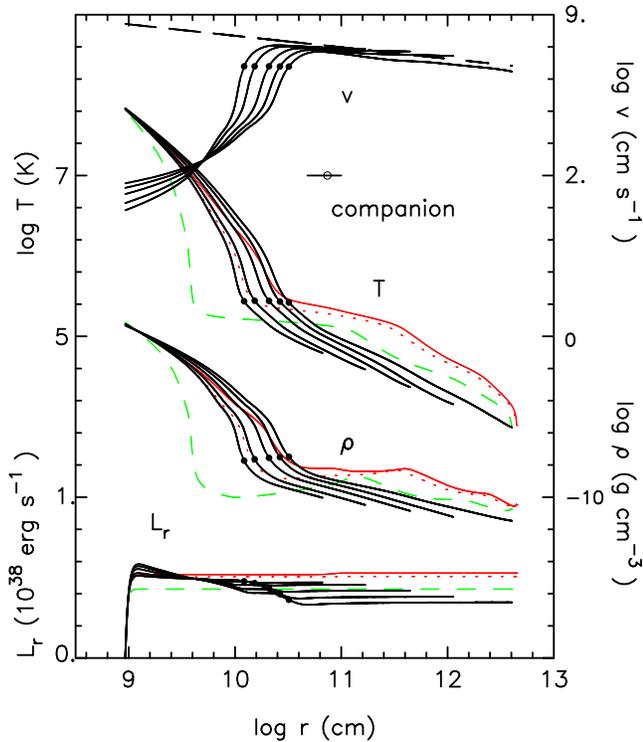}
\caption{
Structure change in the evolution model for V723 Cas. 
The corresponding light curve is shown in Figure \ref{4nova} by the red lines. 
The red solid lines represent the static phase of the first 200 days 
($\log T_{\rm ph}$ (K)=3.892 and $\log R_{\rm ph}$ (cm)=12.65), 
during which the structure hardly changes.
The black solid lines indicates the structure of optically thick wind phase 
in which the photospheric temperature increases with time from  
$\log T_{\rm ph}$ (K)= 3.87, 4.16, 4.37, 4.59, and 4.79. 
The corresponding photospheric radius is 
 $\log R_{\rm ph}$ (cm )= 12,61, 12.05, 11.65, 11.23, 
and 10.82, respectively. 
The dotted red lines indicate a rotating static solution of 
$\eta=0.2$ and 
the green dashed lines the static solutions without the companion effects 
for the same photospheric temperature ($\log T_{\rm ph}$ (K)=3.892) 
as that of the red solid line.
The black dashed line indicates the escape velocity  
$(2GM_{\rm WD}/r)^{1/2}$, where $M_{\rm WD}=0.55~M_\sun$. 
\label{density.evolution}}
\end{figure}

Figure \ref{density.evolution} shows envelope structures
corresponding to each stage of the composite light curve 
model of the 0.55 $M_\sun$ WD (depicted by the red lines
in Figure \ref{4nova}).  The red solid lines in Figure 
\ref{density.evolution} represent the static evolution
in the flat maximum, the structure of which hardly changes
during the flat maximum, and the black solid lines corresponds to 
the wind evolution in the decay phase (after the arrow
in Figure \ref{4nova}).  The green dashed lines indicate the model
without the companion effects. We see that, in the presence of a companion, 
the transition from the static to wind evolution probably occurs,
but may not occur without the companion effects 
because of a large difference between the envelope structures, 
that is, the green dashed line deviates largely from the rightmost
black solid line. 
 
These two trial models adopt a parameter set of the WD mass,
companion mass, chemical composition of the envelope, and the orbit,
which may be close but not the exact values for each nova.
We see, however, our trial light curves reproduce a characteristic
properties of observed light curves; the flat phase of static evolution
followed by a smooth decline due to wind mass loss.  If we choose
another set of the parameters, we can also reproduce the flat peak 
by static solutions and the subsequent decline phase by wind solutions with
a different decay timescale.  The exact parameter fitting is not the aim
of the present work, but we expect that we will find a suitable set
of the parameters which reproduces the light curve as well as 
the other observational constraints.

In the present paper, we studied nova outbursts with a 1D approximation.
The effects of a companion star are, however, highly aspherical, and 
the transition may occur in a different way in the different direction.
This aspherical nature of the transition may relate to aspherical shapes of nova ejecta,  
but far beyond the scope of the present analysis.

\section{DISCUSSION} \label{sec_discussion}

\subsection{Comparison with Other Works} \label{sec_discussion_comparison}

Common envelope evolution with frictional effects by a companion star has been studied 
by multi-dimensional hydrodynamic simulations 
in relation to the formation of close binaries. 
\citet{taa89} showed two-dimensional (2D) evolutions of a common envelope, consisting of 
a 5 $M_{\sun}$ RG and an engulfed 1 $M_{\sun}$  MS star. 
Their results demonstrate that frictional process is strong enough to drive 
a rapid mass outflow from the equatorial plane. Eventually a large part of the envelope
(namely $> 3 M_{\sun}$) would be ejected and the companion would spiral in. 
A similar calculation but for a less massive 2 $M_{\sun}$ RG and the same 1 $M_{\sun}$ MS 
showed that frictional energy is large enough to eject the envelope, but the 
companion does not spiral in \citep{taa91}. 
Three-dimensional (3D) hydrodynamic simulations by \citet{ric08} 
for a binary of a 1.05 $M_{\sun}$ RG 
and an 0.6 $M_{\sun}$ MS companion  
showed that the companion orbit is much 
reduced by frictional effects of the massive 0.69  $M_{\sun}$ RG envelope. 
In these calculations, the companion efficiently interacts with the massive envelope. 
This can be understood from Equation (\ref{equ_drag_simple}).
As the drag luminosity proportionally increases with the density, 
we expect very large drag luminosities in massive envelopes.

\citet{mac80} calculated nova outbursts on a 1.0 $M_{\sun}$ 
WD with a $1.39 \times 10^{-4} M_{\sun}$ envelope 
and an 0.46 $M_{\sun}$ MS companion. The author includes the drag luminosity in 
1D approximation, essentially the same as the present 
work, but the companion's gravity is not included and the opacity is Kramers' law.  
The author found that the expansion velocity reaches $\sim$ 300 km s$^{-1}$,  
whereas 40 km s$^{-1}$ in the case without drag.   
The nova duration is not drastically shortened, i.e., the drag luminosity is 
not effective in mass ejection.

The main difference between ours and MacDonald's (1980) is the opacity. 
With the OPAL opacity we have obtained the strong optically thick winds 
of high wind velocity ($\sim 1000$ km~s$^{-1}$) for a 1.0 $M_{\sun}$ WD   
and a short nova duration ($\sim 1$ yr) as observed. 
Due to the low density at the orbit, the drag luminosity is very small   
\citep{kat94h}. From these reasons, we regard that MacDonald's (1980) results 
are not inconsistent with ours.

\citet{liv90} presented 2D calculations  
of a common envelope phase of a classical nova 
on a 1.0 $M_{\sun}$ WD with a 0.5 $M_{\sun}$ companion and  
with an initially spherical envelope of mass $5 \times 10^{-6}~M_{\sun}$. 
Their calculation showed rapid mass outflows concentrated 
in the orbital plane. 
The mass loss rate reaches $1 \times 10^{-6}~M_{\sun}$ yr$^{-1}$,  
corresponding to a timescale of 5 yr for mass ejection, but still being much 
longer than the observed timescales of novae. 

As already pointed out in our previous work \citep{kat91b}, 
\citet{liv90} assumed adiabatic gas, in which the drag energy is effectively   
consumed to push matter upward against the gravity, and results in the 
strong acceleration of matter. 
If we adopt non-adiabatic gas, all the drag energy deposition 
can escape by diffusion process, as already shown
in Section \ref{sec_results}.  
 Moreover, the authors adopted $60 \times 40 $ grid points in which 
 frictional energy and angular momentum are deposited into four zones near the 
 companion, which may be insufficient to follow the envelope evolution. 
Therefore, we suggest that the adiabatic gas and low-resolution grid may 
be the reason why the authors get strong acceleration of matter concentrated 
in the equatorial plane while the drag luminosity is very small.

After these works done by \citet{mac80} and \citet{liv90}, 
the OPAL opacity appeared which causes 
strong acceleration of nova envelopes. We obtained large mass-loss rates 
and wind velocities as observed \citep{kat94h}. Once the optically thick winds 
occur, the drag luminosity has little effects
\citep[see Section \ref{sec_results} and 
also Section 5 of][]{kat94h}.

To summarize, we may conclude that the drag luminosity is
effective in a massive envelope,
but in a much less massive envelope,
like realistic models of nova outbursts, the drag
luminosity is entirely negligible. 
In other words, the common envelope evolution does not work 
in nova outbursts even if the companion is deeply embedded in the envelope.

\subsection{Spherical (1D) Approximation}
\label{discussion_spherical}

We assumed spherical (1D) approximation of the drag energy deposition,
in which 
the energy is deposited in a shell of radius $R_{\rm orb}$ and thickness of 
$2R_{\rm a}$. In 2D approximation, the energy is deposited in a torus   
of diameter of $2 R_{\rm orb}$ with cross section of $\pi R_{\rm a}^2$. 
Thus, the energy deposition per solid angle may be  
\begin{equation}
{{4 \pi R_{\rm orb}^2 \times 2 R_{\rm a}} \over {2 \pi R_{\rm orb} \times \pi R_{\rm a}^2}}
=2.8  
\end{equation}
\noindent
times larger than in our 1D approximation for the case of
0.55 $M_\sun$ WD and 0.4 $M_\sun$ MS companion.
Here we use the solution of $\log T_{\rm ph}$ (K) =3.85 (the model denoted 
by the black solid lines in Figure \ref{struc.Mcomp}), in which 
$R_{\rm orb}=7.4 \times 10^{10}$ cm and $R_{\rm a}=3.3 \times 10^{10}$ cm. 
In order to simulate this aspherical effect, we calculated a solution 
with enhanced drag luminosity by a factor of 2.8.   
That is, we multiply the right-hand side of Equation (\ref{equ_drag})
by the factor of 2.8. The resultant solution 
 is shown in Figure \ref{struc.Mcomp} by dashed lines. 
The increase of the drag luminosity causes increase of the radiation pressure, 
which results in the density decrease around the orbit. 
Therefore, the resultant drag luminosity is $3.1 \times 10^{36}$ erg~s$^{-1}$, 
only 1.9 times the original one, $1.7 \times 10^{36}$ erg~s$^{-1}$. 
Even if we further increase this factor to 10 as an extreme case,  
the drag luminosity deposition increases only by 3.4 times of the original one, 
to $5.7 \times 10^{36}$ erg~s$^{-1}$. 
With such enhanced energy deposition, the envelope structure changes, but this 
effects are relatively small, i.e., less than the change of
the companion mass of 
0.1 $M_\sun$ as shown in Figure \ref{struc.Mcomp}. In actual multidimensional case,
the energy deposition due to frictional 
effects may escape by diffusion process 
into the pole (rotation-axis) directions, which makes the effects further smaller. 
Therefore, our 1D approximation may not be far from the multi-dimensional
approximations.  

We also assumed that the companion mass is distributed in a spherical shell 
around the orbit.  In the extended envelope, the photospheric radius
is several tens times larger than the companion orbit as seen in Figure \ref{struc.Mcomp}.
Thus, the 
1D approximation may not be so bad in the outer part of the envelope. 
In the vicinity of the orbit, 2D approximation, this may be close
to a time average of a 3D case, gives a stronger gravity 
which may be roughly estimated as 

\begin{equation}
{{4 \pi R_{\rm orb}^2 \times 2 R_{\rm comp}}
\over {2 \pi R_{\rm orb}\times \pi R_{\rm comp}^2}}
=3.6, 
\end{equation}
where $R_{\rm comp}=2.6 \times 10^{10}$ cm. 
Therefore, the companion gravity is much effective than our 1D case. 
The companion masses of V723 Cas, HR Del, and V5558 Sgr are not known,  
but if it is as low as $\sim 0.1-0.2~M_\sun$, as in many classical novae, the 
enhanced gravity may roughly correspond to our massive model of
$0.4 ~M_\sun$.  
Thus, envelope structure may not be largely different from that of
our  1D model.

It should be however noted that the center of rotation is the center of mass,
which is different from the position of the WD.
In this sense, the center of our 1D approximation (i.e., $r=0$)
switches from the WD center in the inner part to the center of mass
in the outer part.  The intermediate region between them is an
interacting region with the companion, which is highly aspherical.
Therefore, our 1D model is a rather crude approximation to this 
intermediate region, which is not easily justified by our 1D model.
We need a full 3-D hydrodynamic simulation including radiative diffusion
to obtain a definite conclusion to the transition.

\subsection{Rotation Law}
\label{discussion_rotation_law}


We also assumed a simple form of angular momentum distribution in the extended 
envelope. Here we examine how this approximation affects the 
envelope structure, assuming a different rotation law. 
Figure \ref{drag} shows a trial angular velocity distribution 
(by the orange solid line) 
that represents co-rotation with the WD, i.e., $\omega=\Omega$, in 
the inner part of the companion ($r < R_{\rm orb}- R_{\rm comp}$, i.e., $\log r$ (cm ) 
$< 10.68$),  
and specific angular momentum conservation in the outside of  
the binary ($r > 3 R_{\rm orb}$, i.e., $\log r$ (cm ) $> 11.35$).
In the intermediate region we connect 
these two laws in an arbitrary form as shown in the figure. 
If the nova envelope expands very slowly, it is 
spun up by the interaction with the companion and gets angular momentum 
$\sim 1.2$--1.7$R_{\rm orb}^2 \Omega$ per unit mass 
\citep[see numerical simulations, e.g.][]{saw84, jah05}. 
This may be the maximum value for actual nova outbursts, 
because expanding speed is not infinitely small    
(in V723 Cas, the rising timescale is one week).   
If no further acceleration occurs outside, the angular velocity decreases 
due to the specific angular momentum conservation. 
We here assume that the angular velocity decreases 
according to the specific angular momentum conservation, 
$l = r^2 \omega \equiv 1.7  R_{\rm orb}^2 \Omega$ at $r > 3R_{\rm orb}$.

Figure \ref{drag} also shows effective mass $fM_{\rm r}$ corresponding to the trial  
angular momentum distribution (the dash-three-dotted line).     
This effective mass $fM_{\rm r}$ is constant ($ =M_{\rm r}$) at $\log r$ (cm ) $< 10.2$ 
as in the other models because the second term of Equation (\ref{equ_f_def}) is small, and we have $f=1$.  
This means that the angular momentum at the WD surface is so small, that the 
results are independent of the initial rotation law of the WD envelope. 
On the other hand, the effective mass $fM_{\rm r}$ varies corresponding to the change 
of $\omega$ in the middle part and quickly rises as $\omega$ decreases 
in the outer part ($\log r$ (cm) $> 11.3$).

Figure \ref{struc.compari} shows  the envelope structure corresponding to the   
trial rotation law by the orange dash-three-dotted lines. 
The distributions of the temperature and density in the inner part 
are very close to those of rapid rotation ($\eta =$ 0.383 and 0.45: 
red solid and purple dash-dotted lines).
As explained previously, a reduced effective gravity due to centrifugal force 
causes smaller Eddington luminosity in the outer part, and thus,  
a smaller diffusive luminosity. 
Therefore, the envelope mass redistributes to balance with a wider local super-Eddington 
region. As a result, the envelope structure becomes closer to 
the one without companion, although   
the structure is very different in the outer part 
due to irregular variation of $fM_{\rm r}$.  

From these results, we say that our rotation law defined in 
Section \ref{centrifugal_force} may be too simple but still useful in a qualitative study of 
possible transition from static to wind evolution of slow novae.

\section{SUMMARY} \label{sec_summary}

Our main results are summarized as follows.

1. We present an idea of a transition from static evolution to 
optically thick wind evolution in low-mass WDs ($\sim0.6~M_\sun$).
In close binaries, if the effects of a companion are included, 
the structure of a static envelope becomes close to that of a wind solution.  
This makes it easy to trigger the transition. 
In wide binaries, the effects of a companion are not important, and  
the structure is very different from that of the wind solution. 
This makes the transition difficult. 

2. A transition from static to wind evolution 
could occur in slow novae such as V723 Cas, HR Del, and V5558 Sgr. 
The transition can explain the spectral evolutions of these novae as 
well as their characteristic light curves. Violent activities like oscillatory behaviors  
in optical light curves are regarded as some relaxation processes associated 
with the transition.   
The presence of a companion deep inside the nova envelopes triggers this transition, which is 
consistent with the orbital separations in short-period binaries like 
V723 Cas. On the other hand, no transition occurs in long-period binaries like PU Vul because of no presence of a companion in the nova envelopes.  

3. We have calculated the drag luminosity for individual novae/helium novae with 
various WD masses. The drag luminosity is negligibly small during the outbursts.




\acknowledgments

We thank the anonymous referee for useful and valuable comments that 
improved the manuscript. 
We also thank the American Association of Variable Star Observers (AAVSO) 
for the photometric data of V723 Cas, HR Del, and V5558 Sgr. 
This research has been supported in part by the Grant-in-Aid for
Scientific Research (20540227, 22540254)
of the Japan Society for the Promotion of Science.

\end{document}